\documentclass[twocolumn]{aastex631}

\shorttitle{Press--Schechter PBH Mass Distributions}
\shortauthors{Farooq et al.}
\submitjournal{ApJL}

\usepackage{amsmath,amssymb}
\usepackage{microtype}

\newcommand{\dd}{\mathrm{d}}
\newcommand{\ii}{\mathrm{i}}
\newcommand{\erfc}{\mathrm{erfc}}

\begin{document}

\title{Press--Schechter Formalism and Primordial Black Hole Mass Distributions}

\correspondingauthor{Owais Farooq}
\email{owai24831@outlook.com}

\author[0000-0000-0000-0000]{Owais Farooq}
\affiliation{Department of Physics, Central University of Kashmir, Ganderbal 191311, India}

\author{Romana Zahoor}
\affiliation{Department of Physics, Central University of Kashmir, Ganderbal 191311, India }

\author{Balungi Francis}
\affiliation{Department of Physics, (Makerere University Kampala, Uganda)}

\begin{abstract}
Primordial black holes (PBHs) can form during radiation domination from rare primordial perturbations that re-enter the Hubble radius and undergo gravitational collapse.
We derive PBH mass distributions using Press--Schechter theory completed by the excursion-set first-crossing construction.
We define the smoothed density contrast $\delta_R$ and its variance $S(R)=\sigma^2(R)$, and connect $S$ to the primordial curvature spectrum $\mathcal{P}_{\mathcal R}(k)$ through the radiation-era transfer.
For Gaussian statistics and a constant collapse threshold $\delta_c$, the formation fraction is an $\erfc$ tail with a controlled rare-event asymptotic.
For a sharp-$k$ filter, $\delta(S)$ is Markovian; solving the diffusion equation with an absorbing barrier yields the first-crossing density
$f(S)=\frac{\delta_c}{\sqrt{2\pi}}S^{-3/2}\exp\!\big(-\delta_c^2/(2S)\big)$.
This gives a differential formation fraction $\dd\beta/\dd\ln M=f(S)\,\big|\dd S/\dd\ln M\big|$ and a mass-conserving formation-era mass function $\dd n_{\rm PBH}/\dd M$.
We then map to the present-day PBH dark-matter fraction per logarithmic mass, $f_{\rm PBH}(M)$, using horizon-entry scaling $M\propto k^{-2}$ and radiation-era redshifting.

\end{abstract}


\section{Introduction}
Primordial black holes (PBHs) provide a link between small-scale primordial fluctuations and late-time probes of compact dark matter.
A predictive PBH calculation requires a formation fraction and a mass distribution.
Press--Schechter theory expresses collapse statistics in terms of smoothed overdensity tails \citep{pressschechter1974}.
Excursion-set theory supplies a first-crossing construction that organizes the smoothing hierarchy \citep{bond1991}.
In the PBH context, careful treatments of the collapse fraction and the mass function within Press--Schechter and excursion-set formalisms have been developed and compared in the literature \citep{green2004new,young2014calculating,sureda2021press,suyama2020novel}.

This work presents a self-contained excursion-set derivation, a transparent mapping from a curvature spectrum $\mathcal{P}_{\mathcal R}(k)$ to a PBH mass distribution, and a concise constraint-ready representation in terms of the present-day fraction per logarithmic mass, $f_{\rm PBH}(M)$.
The baseline assumptions are: radiation domination at formation, Gaussian statistics for the linear density contrast at horizon entry, a constant threshold $\delta_c$, and a sharp-$k$ filter that yields Markovian walks in variance time.
Threshold systematics and profile dependence are anchored to numerical-relativity results \citep{harada2013,musco2019}.
Extensions include peak statistics, nonlinear overdensity statistics, and non-Gaussian corrections that modify collapse tails and inferred mass functions \citep{wu2020peak,wang2021perturbations,mahbub2020impact,pi2024nonGaussianities,germani2023statistics}.
\vspace{2.5cm}
\section{Press--Schechter and excursion-set derivation}
\subsection{Smoothing and variance.}
Let $\delta(\mathbf{x})\equiv(\rho(\mathbf{x})-\bar\rho)/\bar\rho$ be the density contrast.
With Fourier conventions
\begin{equation}
\begin{split}
\delta(\mathbf{x})&=\int\frac{\dd^3k}{(2\pi)^3}\,\delta(\mathbf{k})\,e^{\ii\mathbf{k}\cdot\mathbf{x}},\\
\langle\delta(\mathbf{k})\,\delta^*(\mathbf{k}')\rangle&=(2\pi)^3\,\delta^{(3)}(\mathbf{k}-\mathbf{k}')\,P_\delta(k),
\end{split}
\end{equation}
define the smoothed field at scale $R$ by
\begin{equation}
\delta_R(\mathbf{x})=\int\frac{\dd^3k}{(2\pi)^3}\,\delta(\mathbf{k})\,W(kR)\,e^{\ii\mathbf{k}\cdot\mathbf{x}}.
\end{equation}
The variance is
\begin{equation}
\begin{split}
S(R)\equiv\sigma^2(R)&=\langle\delta_R^2\rangle=\int_0^\infty\frac{\dd k}{k}\ \mathcal{P}_\delta(k)\,W^2(kR),\\
\mathcal{P}_\delta(k)&=\frac{k^3}{2\pi^2}P_\delta(k).
\label{eq:S_def}
\end{split}
\end{equation}

\subsection{Radiation-era transfer to curvature power.}
During radiation domination, a standard relation between comoving-gauge density contrast and curvature perturbation is
\begin{equation}
\delta(\mathbf{k},\eta)=\frac{4}{9}\left(\frac{k}{aH}\right)^2 T(k\eta)\,\mathcal{R}(\mathbf{k}),
\label{eq:delta_transfer}
\end{equation}
with $T(1)=\mathcal{O}(1)$ near horizon entry.
For the variance at smoothing scale \(R\), \eqref{eq:delta_transfer} is evaluated at the horizon-entry epoch associated with that scale, \(aH=R^{-1}\), giving \(q/(aH)=qR\) and \(q\eta=qR\) in \eqref{eq:sigma_PR}.
This gives the coarse-grained variance in terms of the curvature spectrum $\mathcal{P}_{\mathcal R}$:
\begin{equation}
\sigma^2(R)=\frac{16}{81}\int_0^\infty\frac{\dd q}{q}\,(qR)^4\,T^2(qR)\,\mathcal{P}_{\mathcal R}(q)\,W^2(qR).
\label{eq:sigma_PR}
\end{equation}

\subsection{Gaussian collapse tail.}
Assume $\delta_R$ is Gaussian with mean $0$ and variance $S(M)$.
For a constant collapse threshold $\delta_c$, the Press--Schechter tail fraction is
\begin{equation}
\begin{split}
\beta_{\rm tail}(M)&=\int_{\delta_c}^{\infty}\frac{1}{\sqrt{2\pi S(M)}}\exp\!\left(-\frac{\delta^2}{2S(M)}\right)\dd\delta \\
&=\frac{1}{2}\erfc\!\left(\frac{\delta_c}{\sqrt{2S(M)}}\right).
\label{eq:beta_tail}
\end{split}
\end{equation}
In the sharp-$k$ excursion-set construction, the formation fraction is identified with the crossing probability,
\begin{equation}
\beta(M)\equiv P_{\rm cross}(S(M))=\erfc\!\left(\frac{\delta_c}{\sqrt{2S(M)}}\right)=2\,\beta_{\rm tail}(M).
\label{eq:beta_cross_def}
\end{equation}
With $\nu(M)=\delta_c/\sqrt{S(M)}$, the rare-event regime $\nu\gg 1$ yields
\begin{equation}
\beta_{\rm tail}(M)\simeq \frac{1}{\nu(M)\sqrt{2\pi}}\exp\!\left(-\frac{\nu^2(M)}{2}\right).
\label{eq:beta_asymp}
\end{equation}

\subsection{Excursion-set first crossing for sharp-$k$ filter.}
Take the sharp-$k$ filter $W(kR)=\Theta(1-kR)$.
Then $\delta(S)$ is a Markov random walk with increments $\delta(S+\Delta S)-\delta(S)\sim \mathcal{N}(0,\Delta S)$ \citep{bond1991,sureda2021press}.
Let $\Pi(\delta,S)$ be the density of walks at value $\delta$ and variance time $S$ subject to absorbing collapse at $\delta_c$:
\begin{equation}
\frac{\partial\Pi}{\partial S}=\frac{1}{2}\frac{\partial^2\Pi}{\partial\delta^2},\qquad
\Pi(\delta,0)=\delta_{\rm D}(\delta),\qquad
\Pi(\delta_c,S)=0.
\label{eq:diffusion}
\end{equation}
The method of images gives
\begin{equation}
\Pi(\delta,S)=\frac{1}{\sqrt{2\pi S}}
\left[
\exp\!\left(-\frac{\delta^2}{2S}\right)-\exp\!\left(-\frac{(2\delta_c-\delta)^2}{2S}\right)
\right].
\label{eq:images}
\end{equation}
The survival probability is $F(S)=\int_{-\infty}^{\delta_c}\Pi(\delta,S)\dd\delta=\mathrm{erf}\!\left(\delta_c/\sqrt{2S}\right)$.
Hence the crossing probability is $P_{\rm cross}(S)=1-F(S)=\erfc\!\left(\delta_c/\sqrt{2S}\right)$ and the first-crossing density is
\begin{equation}
\begin{split}
f(S)=\frac{\dd}{\dd S}P_{\rm cross}(S)&=\frac{\delta_c}{\sqrt{2\pi}}\,S^{-3/2}\exp\!\left(-\frac{\delta_c^2}{2S}\right),
\\
\int_0^\infty f(S)\dd S&=1.
\label{eq:fS}
\end{split}
\end{equation}

\subsection{Differential formation fraction and formation-era mass function.}
With $S=S(M)$, define
\begin{equation}
\frac{\dd\beta}{\dd\ln M}=f(S)\,\left|\frac{\dd S}{\dd\ln M}\right|=f(S)\,M\left|\frac{\dd S}{\dd M}\right|.
\label{eq:dbeta_dlnM}
\end{equation}
Mass conservation at formation time $t_f$ yields
\begin{equation}
\frac{\dd n_{\rm PBH}}{\dd M}(t_f)=\frac{\rho_{\rm tot}(t_f)}{M^2}\,\frac{\dd\beta}{\dd\ln M}.
\label{eq:dndM_form}
\end{equation}

\section{Mass mapping and present-day fraction}
\label{sec:mass_mapping}

\subsection{Horizon entry, smoothing scale, and the PBH mass scale}
A formation-scale comoving mode \(k\) is associated with horizon entry at
\begin{equation}
k = a_f H_f ,
\label{eq:k_hor_entry}
\end{equation}
where the subscript \(f\) denotes evaluation at the PBH formation epoch during radiation domination.
A smoothing scale \(R\) that tracks the comoving mode is taken as \(R\simeq k^{-1}\), so that the sharp-\(k\) filter \(W(kR)=\Theta(1-kR)\) selects modes \(q\le R^{-1}\).

The horizon mass at formation is
\begin{equation}
M_H(t_f)\equiv \frac{4\pi}{3}\,\rho_f\,H_f^{-3}
= \frac{1}{2G}\,H_f^{-1},
\label{eq:MH_def}
\end{equation}
using \(\rho_f=3H_f^2/(8\pi G)\).
The PBH mass is modeled as a fraction \(\gamma\) of the horizon mass,
\begin{equation}
M \equiv \gamma\, M_H(t_f)=\frac{\gamma}{2G}\,H_f^{-1},
\qquad \gamma=\mathcal{O}(0.1\text{--}1),
\label{eq:M_gamma_MH}
\end{equation}
with \(\gamma\) encoding collapse efficiency and profile dependence \citep{harada2013,musco2019,carr2016primordial,sureda2021press}.

Radiation domination gives a temperature-based Hubble scale,
\begin{equation}
H_f \simeq 1.66\,g_*^{1/2}(T_f)\,\frac{T_f^2}{M_{\rm Pl}},
\label{eq:H_T_RD}
\end{equation}
and entropy conservation gives
\begin{equation}
a_f \propto g_{*s}^{-1/3}(T_f)\,T_f^{-1},
\label{eq:a_T_entropy}
\end{equation}
with \(g_*(T)\) and \(g_{*s}(T)\) the energy and entropy effective relativistic degrees of freedom.
Combining \eqref{eq:k_hor_entry}--\eqref{eq:a_T_entropy} yields the characteristic scaling
\begin{equation}
\begin{split}
M \propto H_f^{-1} \propto T_f^{-2},&
\qquad
k \propto a_f H_f \propto g_*^{1/2} g_{*s}^{-1/3} T_f
\\
&\Rightarrow\quad M \propto k^{-2}
\label{eq:M_k_scaling}
\end{split}
\end{equation}
up to mild \(g_*\) and \(g_{*s}\) dependence.
A widely used numerical mapping \citep{carr2016primordial,sureda2021press} is
\begin{equation}
M(k)\simeq 10^{18}\,{\rm g}\,
\left(\frac{\gamma}{0.2}\right)
\left(\frac{g_*(T_f)}{106.75}\right)^{-1/6}
\left(\frac{k}{7\times 10^{13}\,{\rm Mpc}^{-1}}\right)^{-2},
\label{eq:M_of_k_expanded}
\end{equation}
equivalent to \(M\propto k^{-2}\) with the radiation-era \(g_*\) correction.

The Jacobians needed for converting \(k\)-space to mass space follow directly from \eqref{eq:M_k_scaling}:
\begin{equation}
\frac{\dd \ln M}{\dd \ln k}=-2,
\qquad
\frac{\dd \ln k}{\dd \ln M}=-\frac12,
\qquad
\frac{\dd \ln R}{\dd \ln M}=+\frac12,
\label{eq:jacobians_Mk}
\end{equation}
where \(R\simeq k^{-1}\).

\subsection{Sharp-\texorpdfstring{\(k\)}{k} variance and an explicit \texorpdfstring{\(\dd S/\dd\ln M\)}{dS/dlnM}}
For the sharp-\(k\) window \(W(qR)=\Theta(1-qR)\), the variance \eqref{eq:S_def} becomes
\begin{equation}
S(R)=\sigma^2(R)
=\int_0^{1/R}\frac{\dd q}{q}\,\mathcal{P}_\delta(q).
\label{eq:S_sharpk}
\end{equation}
Differentiation with respect to \(\ln R\) yields a boundary-local relation,
\begin{equation}
\frac{\dd S}{\dd \ln R}
= -\,\mathcal{P}_\delta\!\left(\frac{1}{R}\right),
\label{eq:dS_dlnR_sharpk}
\end{equation}
and using \eqref{eq:jacobians_Mk} gives
\begin{equation}
\frac{\dd S}{\dd \ln M}
=\frac{\dd S}{\dd \ln R}\,\frac{\dd \ln R}{\dd \ln M}
= -\frac12\,\mathcal{P}_\delta\!\left(k\right),
\qquad (k=R^{-1}).
\label{eq:dS_dlnM_sharpk}
\end{equation}
In applications, the differential mass fraction uses the positive measure \(\dd S\) along the excursion-set trajectory, so the mass-space density is taken with the magnitude of the Jacobian.

\subsection{From first crossing to a present-day fraction per logarithmic mass}
The excursion-set first-crossing density \(f(S)\) in \eqref{eq:fS} yields the differential collapse fraction at formation,
\begin{equation}
\frac{\dd\beta}{\dd\ln M}
= f(S)\,\left|\frac{\dd S}{\dd\ln M}\right|
= f\!\big(S(M)\big)\,\left|\frac{\dd S}{\dd\ln M}\right|.
\label{eq:dbeta_dlnM_abs}
\end{equation}
The total collapse fraction is \(\beta_{\rm tot}=\int (\dd\beta/\dd\ln M)\,\dd\ln M\).
For a narrow (effectively monochromatic) population centered at \(M\), one may identify \(\beta(M)\) with the integrated fraction in that bin.

The mapping from formation fraction to a present-day PBH dark-matter fraction follows from redshifting.
At formation during radiation domination,
\begin{equation}
\rho_{\rm PBH}(t_f) = \beta_{\rm tot}\,\rho_{\rm tot}(t_f)\simeq \beta_{\rm tot}\,\rho_{\rm rad}(t_f),
\label{eq:rhoPBH_form}
\end{equation}
and during radiation domination \(\rho_{\rm PBH}\propto a^{-3}\) while \(\rho_{\rm rad}\propto a^{-4}\), giving growth of the PBH-to-radiation ratio proportional to \(a\).
At matter--radiation equality \(t_{\rm eq}\),
\begin{equation}
\left.\frac{\rho_{\rm PBH}}{\rho_{\rm rad}}\right|_{\rm eq}
=\beta_{\rm tot}\,\frac{a_{\rm eq}}{a_f}.
\label{eq:ratio_eq}
\end{equation}
For \(t\ge t_{\rm eq}\) the ratio \(\rho_{\rm PBH}/\rho_{\rm m}\) becomes constant since both scale as \(a^{-3}\).
Defining the present-day fraction per logarithmic mass,
\begin{equation}
\begin{split}
f_{\rm PBH}(M)&\equiv \frac{1}{\Omega_{\rm DM}}\frac{\dd\Omega_{\rm PBH,0}}{\dd\ln M},
\\
f_{\rm PBH}^{\rm tot}&=\int f_{\rm PBH}(M)\,\dd\ln M,
\label{eq:fPBH_def_expanded}
\end{split}
\end{equation}
a standard radiation-era conversion used in PBH phenomenology \citep{young2014calculating,carr2016primordial,sureda2021press,suyama2020novel} is obtained by expressing \(a_{\rm eq}/a_f\) in terms of \(T_f\) and then eliminating \(T_f\) in favor of \(M\) using \eqref{eq:M_gamma_MH}--\eqref{eq:H_T_RD}.
This yields the familiar \(M^{-1/2}\) scaling:
\begin{equation}
\begin{split}
f_{\rm PBH}(M)
\simeq
\mathcal{A}\,
\left(\frac{\gamma}{0.2}\right)^{3/2}
\left(\frac{g_*(T_f)}{106.75}\right)^{-1/4} \\
\cdot \left(\frac{M}{10^{18}\,{\rm g}}\right)^{-1/2}
\left(\frac{\dd\beta}{\dd\ln M}\right),
\label{eq:fPBH_from_dbeta}
\end{split}
\end{equation}
where \(\mathcal{A}\) is a numerical normalization fixed by cosmological parameters and the equality temperature, and the bracketed scaling captures the dominant formation-era dependence.
For a sharply peaked mass function, \(\dd\beta/\dd\ln M\) reduces to \(\beta(M)\) in that narrow bin and \eqref{eq:fPBH_from_dbeta} reduces to the commonly quoted monochromatic mapping, represented in your Equation~\eqref{eq:Omega_from_beta}.
\begin{equation}
\frac{\Omega_{\rm PBH}(M)}{\Omega_{\rm DM}}
\simeq
\left(\frac{\beta(M)}{8\times 10^{-16}}\right)
\left(\frac{\gamma}{0.2}\right)^{3/2}
\left(\frac{g_*(T_f)}{106.75}\right)^{-1/4}
\left(\frac{M}{10^{18}\,{\rm g}}\right)^{-1/2}.
\label{eq:Omega_from_beta}
\end{equation}

The present-day differential number density follows directly from \eqref{eq:fPBH_def_expanded}:
\begin{equation}
\frac{\dd n_{\rm PBH}}{\dd M}\bigg|_{0}
=
\frac{\rho_{c,0}\,\Omega_{\rm DM}}{M^2}\,f_{\rm PBH}(M),
\label{eq:dndM_today}
\end{equation}
where \(\rho_{c,0}\) is the critical density today.
Equations \eqref{eq:sigma_PR}, \eqref{eq:fS}, \eqref{eq:dbeta_dlnM_abs}, and \eqref{eq:fPBH_from_dbeta} provide a closed pipeline from \(\mathcal{P}_{\mathcal R}(k)\) to \(f_{\rm PBH}(M)\).

\section{Constraints and illustration}
\label{sec:constraints}

\subsection{Extended-mass-function use of monochromatic limits}
Many observational probes provide upper envelopes \(f_{\max}(M)\) calibrated under a monochromatic assumption.
A conservative and widely used treatment for an extended distribution is the channel-wise integral criterion \citep{carr2017extended,sureda2021press},
\begin{equation}
\int \frac{f_{\rm PBH}(M)}{f_{\max}(M)}\,\dd\ln M\le 1,
\label{eq:extended_constraint_expanded}
\end{equation}
applied separately to each constraint channel and then combined by requiring satisfaction for every channel considered.
In practice, one evaluates \(f_{\rm PBH}(M)\) from \eqref{eq:fPBH_from_dbeta} and integrates numerically over the mass range where \(f_{\max}(M)\) is provided.

A compact analytic proxy frequently used for illustrative extended distributions is the lognormal form,
\begin{equation}
f_{\rm PBH}(M)=
\frac{f_{\rm PBH}^{\rm tot}}{\sqrt{2\pi}\,\sigma_{\ln M}}
\exp\!\left[
-\frac{\big(\ln(M/M_c)\big)^2}{2\sigma_{\ln M}^2}
\right],
\label{eq:lognormal_fPBH}
\end{equation}
parameterized by a central mass \(M_c\) and width \(\sigma_{\ln M}\).
This proxy is convenient for demonstrating the effect of width on \eqref{eq:extended_constraint_expanded}, and it also offers a useful approximation to spectra that arise from localized features in \(\mathcal{P}_{\mathcal R}(k)\).

\subsection{Representative channels and interpretation}
Microlensing constraints bound compact-object fractions across asteroid-to-stellar mass scales, while CMB anisotropy and spectral-distortion constraints bound accretion and energy-injection effects for heavier PBHs \citep{niikura2019,alihaimoud2017,carr2017extended}.
Given a candidate \(f_{\rm PBH}(M)\), each channel yields a scalar consistency check via \eqref{eq:extended_constraint_expanded}.
When multiple channels are used, an equivalent summary diagnostic is
\begin{equation}
\begin{split}
&\mathcal{I}_{\rm ch}\equiv \int \frac{f_{\rm PBH}(M)}{f_{\max,{\rm ch}}(M)}\,\dd\ln M,
\\
&\text{viability requires}\quad \mathcal{I}_{\rm ch}\le 1\ \ \text{for all channels}.
\label{eq:Ich_def}
\end{split}
\end{equation}
This representation highlights two mass-function effects:
broadening \(f_{\rm PBH}(M)\) typically increases \(\mathcal{I}_{\rm ch}\) by sampling multiple constrained mass decades, and shifting \(M_c\) moves the dominant support between channels.

\begin{figure}
\centering
\includegraphics[width=\columnwidth]{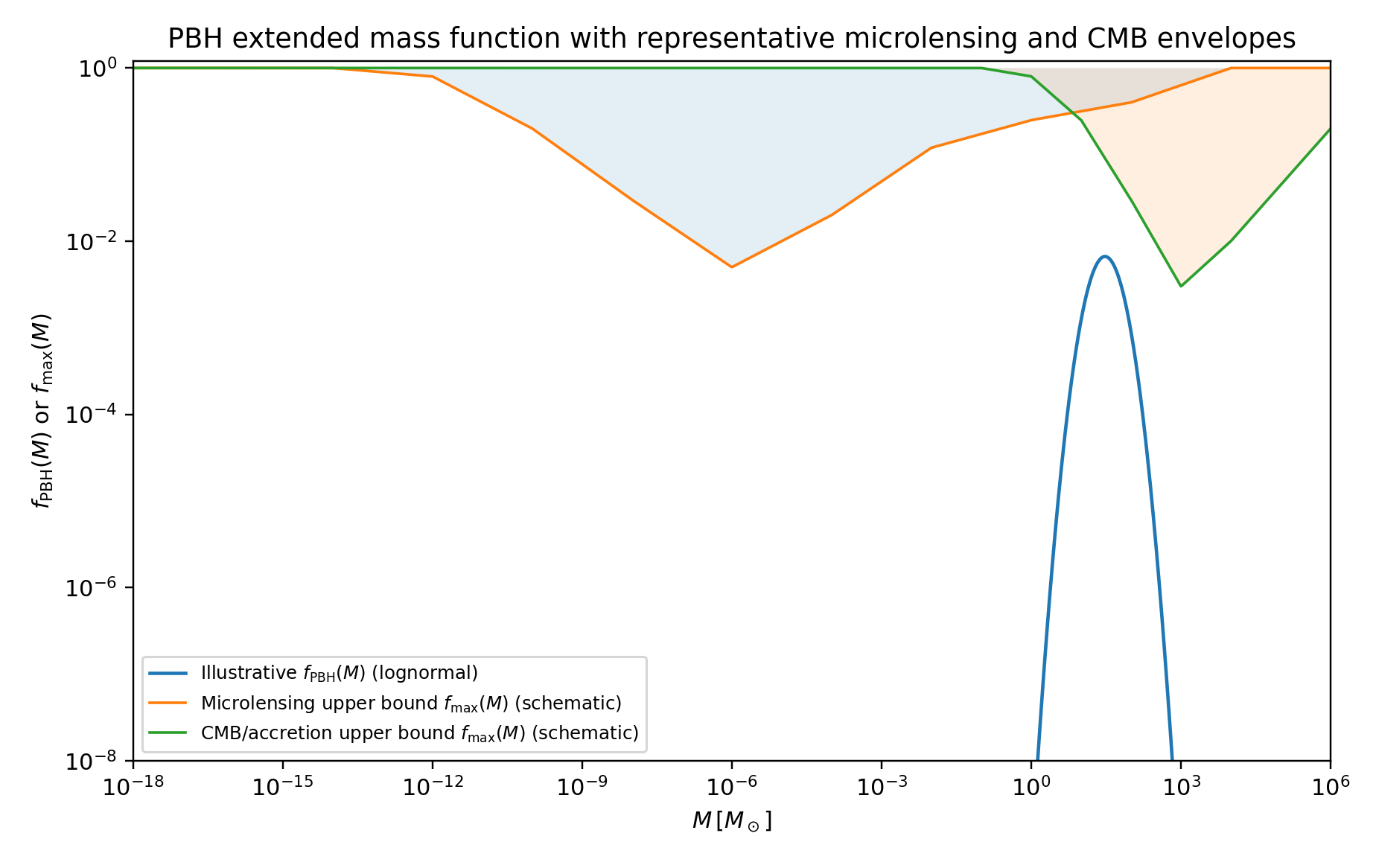}
\caption{Illustrative present-day PBH fraction per logarithmic mass \(f_{\rm PBH}(M)\), shown as a lognormal proxy \eqref{eq:lognormal_fPBH}, over representative microlensing and CMB-accretion envelopes \(f_{\max}(M)\).
For a given observational channel, quantitative extended-distribution consistency is evaluated through the channel-wise integral criterion \eqref{eq:extended_constraint_expanded}.}
\label{fig:pbh_constraints}
\end{figure}

\section{Conclusion}
\label{sec:conclusion}
This work developed a compact, end-to-end derivation of primordial black-hole (PBH) mass distributions within Press--Schechter theory completed by the excursion-set first-crossing construction. Starting from the smoothed density contrast \(\delta_R\) and its variance \(S(R)\), we connected the density power \(\mathcal{P}_\delta(k)\) to the primordial curvature spectrum \(\mathcal{P}_{\mathcal R}(k)\) through the standard radiation-era transfer. For Gaussian statistics and a constant collapse threshold \(\delta_c\), the sharp-\(k\) choice yields Markovian random walks \(\delta(S)\), so the collapse problem reduces to a diffusion equation with an absorbing barrier. The method-of-images solution provides the closed-form first-crossing density \(f(S)\), which directly generates the differential formation fraction \(\dd\beta/\dd\ln M\) and a mass-conserving formation-era mass function \(\dd n_{\rm PBH}/\dd M\).

The mapping from formation scales to present-day abundance was then made explicit through horizon-entry scaling \(M\propto k^{-2}\) during radiation domination and the standard redshifting relation that converts \(\dd\beta/\dd\ln M\) into the present-day PBH dark-matter fraction per logarithmic mass \(f_{\rm PBH}(M)\). This representation interfaces directly with observational constraints through the channel-wise extended-mass-function criterion \(\int f_{\rm PBH}(M)/f_{\max}(M)\,\dd\ln M\le 1\), which provides a practical, constraint-ready summary for any model that predicts \(\mathcal{P}_{\mathcal R}(k)\).

Several theoretically motivated upgrades fit naturally within the same pipeline. A scale-dependent or profile-dependent collapse criterion can be implemented as a moving barrier \(\delta_c=\delta_c(S)\), modifying the first-crossing problem while preserving the excursion-set logic. Window functions other than sharp-\(k\) introduce correlated steps in \(\delta(S)\), leading to non-Markovian first-crossing statistics and calculable corrections to \(f(S)\). Non-Gaussianity in the primordial perturbations enters through modified tail probabilities and altered first-crossing statistics, reshaping \(\dd\beta/\dd\ln M\) and shifting inferred constraints on \(\mathcal{P}_{\mathcal R}(k)\). Critical-collapse scaling, in which the PBH mass depends continuously on the distance above threshold, provides a further route to extended mass functions even for sharply featured power spectra, and it can be incorporated by convolving the first-crossing statistics with the critical-scaling mass relation. Together, these extensions position the present framework as a flexible baseline for translating small-scale primordial physics into PBH mass distributions and observationally testable abundance constraints.

\end{document}